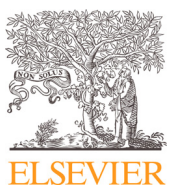
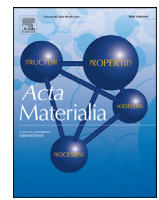

Full length article

# Hexaferrite-based permanent magnets with upper magnetic properties by cold sintering process via a non-aqueous solvent

Aida Serrano [a,*], Eduardo García-Martín [a,b], Cecilia Granados-Miralles [a], Giulio Gorni [c], Jesús López-Sánchez [d], Sandra Ruiz-Gómez [b,c], Lucas Pérez [b], Adrián Quesada [a], José F. Fernández [a]

[a] *Departamento de Electrocerámica, Instituto de Cerámica y Vidrio (ICV), CSIC, Madrid E-28049, Spain*
[b] *Departamento de Física de Materiales, Universidad Complutense de Madrid, Madrid E-28040, Spain*
[c] *CELLS-ALBA Synchrotron, Cerdanyola del Vallès E-08290, Spain*
[d] *Instituto de Magnetismo Aplicado, UCM-ADIF, Las Rozas E-28232, Spain*

## a b s t r a c t

The incessant technological pursuit towards a more sustainable and green future depends strongly on permanent magnets. At present, their use is widespread, making it imperative to develop new processing methods that generate highly competitive magnetic properties reducing the fabrication temperatures and costs. Herein, a novel strategy for developing dense sintered magnets based on Sr-hexaferrites with upper functional characteristics is presented. An innovative cold sintering approach using glacial acetic acid as novelty, followed by a post-annealing at 1100 °C, achieves a densification of the ceramic magnets of 92% with respect to the theoretical density and allows controlling the particle growth. After the cold sintering process, a fraction of amorphous SrO is identified, in addition to a partial transformation to $\alpha$-Fe$_2$O$_3$ as secondary crystalline phase. 46 wt% of SrFe$_{12}$O$_{19}$ remains, which is mostly recuperated after the post-thermal treatment. These findings do not significantly modify the final structure of ferrite magnets, neither at short- nor long-range order. The innovative process has a positive impact on the magnetic properties, yielding competitive ferrite magnets at lower sintering temperatures with an energy efficiency of at least 25%, which opens up a new horizon in the field of rare-earth free permanent magnets and new possibilities in other applications.

## 1. Introduction

Constant technological requirements lead to research into new materials and development of existing ones with new or improved attractive functional properties. Specifically, the M-type hexaferrites (MFe$_{12}$O$_{19}$, M = Pb, Sr, Ba) exhibit remarkable structural and hard magnetic properties [1–3], and they are widely investigated [2,4,5]. Ferrites have a low cost as compared to their competitor permanent magnets (AlNiCo, MnBi, MnAl or NdFeB) and they are oxides. Therefore, they do not present the corrosion problems of metallic alloys-based magnets and their manufacturing process is much easier and cheaper with a stability of operation over time, temperature, and radiation [6]. Besides, the intrinsic characteristics of hexaferrites make them a suitable and green alternative to rare-earth-based permanent magnets in a variety of current applications, such as in small electric motors, microwave devices, recording media, telecommunication and electronic industry, and also in micro and nano-systems such as biomarkers, biodiagnostics and biosensors [7–11].

The ceramic magnets based on hexaferrites usually require high sintering temperatures (> 1200 °C) [12] and long dwell times to achieve density values > 90% with respect to the theoretical density. However, during the solid-state sintering using thermal conventional methodologies, a particle coarsening process towards an exaggerated growth takes place, which is detrimental to the magnetic response [13,14]. Currently, many strategies are being investigated to avoid these issues and the results are often unpromising. For example, the incorporation of secondary phases is required to ensure densification at moderate particle sizes [12] but the sintering temperature are not reduced < 1200 °C.

An interesting approach to fabricate the ceramics magnets at lower temperatures and sintering times, i.e. from a greener perspective, may be the cold sintering process (CSP). The CSP is a rel-atively new sintering route [15] in which inorganic powder compounds are mixed with an aqueous solvent that partially solubilizes, promoting mass transport at low temperatures under an applied uniaxial pressure, obtaining a dense material [16–18]. This sintering process has been employed for the densification of a large number of compounds using aqueous solvents, where many parameters can be considered to tailor the final properties of sintered materials [15–18]. Some works in this matter have even considered the use of pure organic solvents during the CSP. Berbano et al. employed ethanol in the CSP of $Li_{1.5}Al_{0.5}Ge_{1.5}(PO_4)_3$ solid electrolyte [19], Ndayishimiye et al. used TEOS as transient liquid in the preparation of $(1-x)\, SiO_2 - x\, PTFE$ composites [20] and Kang et al. demonstrated the use of dimethyl sulfoxide solutions for the densification of metal oxides [21]. However, despite the great advances made in such a short time, there are still many open questions and materials that have not been considered or sintered by this innovative way. No CSP studies on permanent magnets have been reported on this matter.

* Corresponding author.
  *E-mail address:* aida.serrano@icv.csic.es (A. Serrano).

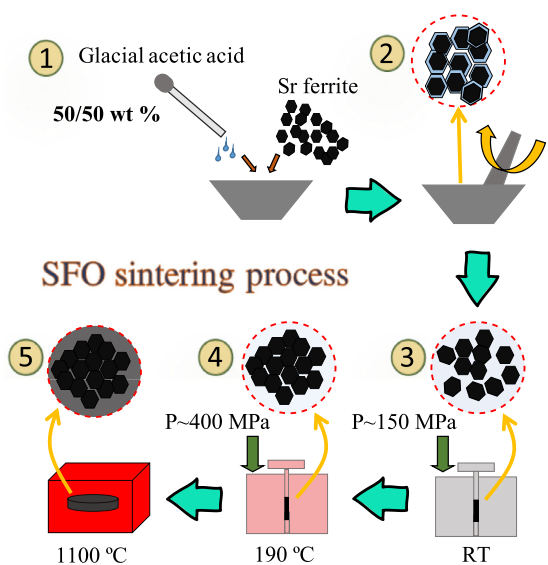

**Fig. 1.** Representative scheme of the cold sintering route using GAA followed by a post-annealing step at 1100 °C for 2 h for obtaining dense SFO ceramics, 1→5.

Here, a novel route based on CSP employing a non-aqueous solvent is followed to sinter dense permanent magnets based on Sr-hexaferrites. The innovative approach allows the structural and compositional control of the ceramic magnets, reaching improved magnetic properties at lower sintering temperatures and with greater energy efficiency. This work also provides a further understanding of the sintering of hexaferrite-based ceramics, which open up a new horizon in the field of rare-earth free permanent magnets and beyond.

## 2. Experimental methods

### 2.1. Materials and processing route

Dense hexaferrite ceramics were reached by CSP using glacial acetic acid (GAA) as a pure organic solvent plus a subsequent post-annealing. Fig. 1 illustrates the sintering steps followed. Prior to the CSP, commercial platelet-shaped $SrFe_{12}O_{19}$ (SFO) particles from Max Baermann GmbH Holding (Germany) and GAA from Sigma-Aldrich were mixed in a proportion of 50/50 wt% in an agate mortar for 10 min until the solvent is distributed homogeneously around the ferrite particles (Steps 1 and 2). The initial excess of GAA ensures its homogeneous distribution and the consequent par-tial dissolution of the particle surfaces. Subsequently, the resulting wetted mixture of SFO plus GAA was pressed at 150 MPa (green piece, labelled G) for 5 min at room temperature (RT) in a cylindrical die (Step 3). The pieces pressed into the die were submitted to CSP (labelled CSP) heating with an annealing rate of 20 °C min$^{-1}$ at 190 °C and applying a pressure of 400 MPa for 2 h in air atmosphere (Step 4). CSP samples were then extracted from the die and thermally treated at 1100 °C for 2 h in air (Step 5) to obtain the final magnets (CSP1100°C). These sintered ceramics were compared with SFO-based magnets obtained by a conventional process, which were pressed at 270 MPa and subsequently sintered at 1100 °C (Cn1100°C) and 1300 °C (Cn1300°C) for 4 h in air.

### 2.2. Characterisation procedure

The relative density of the SFO-based pieces was evaluated by the Archimedes method and by the mass/dimensions values. Relative densities were calculated considering the theoretical density of SFO (5.10 g cm$^{-3}$) [22], $\alpha$-$Fe_2O_3$ (5.26 g cm$^{-3}$) [23] and SrO (5.01 g cm$^{-3}$) [24], identified by X-ray diffraction (XRD) and X-ray absorption spectroscopy (XAS) techniques.

The morphology of the SFO ceramics was analyzed by field emission scanning electron microscopy (FESEM), with an S-4700 Hitachi instrument at 20 kV. Samples were studied on fresh fractured surfaces, except to those sintered by conventional route that were polished and thermally etched at a temperature value of 10% below the sintering temperature. ImageJ software was employed to generate the particle size distribution from FESEM images and determine the average particle size.

XRD measurements were carried out in a D8 Advanced Bruker diffractometer using a Lynx Eye detector and a Cu K$\alpha$ radiation with a $\lambda = 0.154$ nm in the $2\theta$ range 25–65 deg. Rietveld refinements of the XRD data were carried out using the software FullProf [25] in order to obtain quantitative information from the samples. The propagation of errors is considered to calculate the uncertainty in the refined parameters. In the refinements, a Thompson-Cox-Hastings pseudo-Voigt function was used to model the peak-profile [26]. The instrumental-contribution to the peak-broadening was determined from measurements of a standard powder (NIST LaB6 SRM® 660b) [27] and deconvoluted from the data. The sample-contribution to the broadening was considered as purely size-originated.

XAS was performed on the sintered samples to evaluate the effect of the sintering method in the short-range structural modifications of the SFO structure. Powdered samples were characterized at the Fe and Sr K-edge at the BL22 CLÆSS beamline of the ALBA synchrotron facility (Barcelona, Spain). XAS measurements were collected on pellets prepared by mixing proper amount of sample with cellulose, at RT in transmission mode. The monochromator used in the experiments was a double Si crystal oriented in the (311) direction, with an incident energy resolution less than 0.3 eV at Fe K-edge. Higher harmonic rejection was achieved by a proper choice of angle and coating of collimating and focusing mirrors. The incident and transmitted intensity were detected with two ionization chambers filled with $N_2$ and inert gases in order to have 10 and 75% absorption, respectively. Metal foils were measured and used as reference to calibrate the energy. SFO starting powders, $\alpha$-$Fe_2O_3$ and SrO powder references were also measured as reference from transmitted photons.

For the theoretical extended X-ray absorption fine structure (EXAFS) calculations, Fourier transform was performed in the $k^3\chi(k)$ weighted EXAFS signal between 2.5 and 13.0 Å$^{-1}$ at the Fe K-edge and between 3 and 11.5 Å$^{-1}$ at the Sr K-edge. Experimental EXAFS results were fitted in $R$-space in the range 1.4–3.8 Å and 2.5–4.4 Å at the Fe and Sr K-edge, respectively, using the FEFFIT code [28,29]. The fitting was performed by fixing the shift at the edge energy for each absorption edge, which were previously calculated from the starting SFO powders (used as reference). Therefore, the coordination number $N$, the interatomic distance $R$ and the Debye-Waller (DW) factors for each shell are used as free parameters. Only the most intense single-scattering paths were considered for the fitting. At the Fe K-edge position, a first shell produced by the interaction of a Fe absorbing atom with six O atoms and a second one constituted by two shells of six Fe atoms each were considered. At the Sr K-edge position, a three-shell model was considered with a first and a second shell produced each by the interaction of a Sr absorbing atom with six O atoms and a third shell formed by the interaction of a Sr absorbing atom with fifteen Fe atoms.

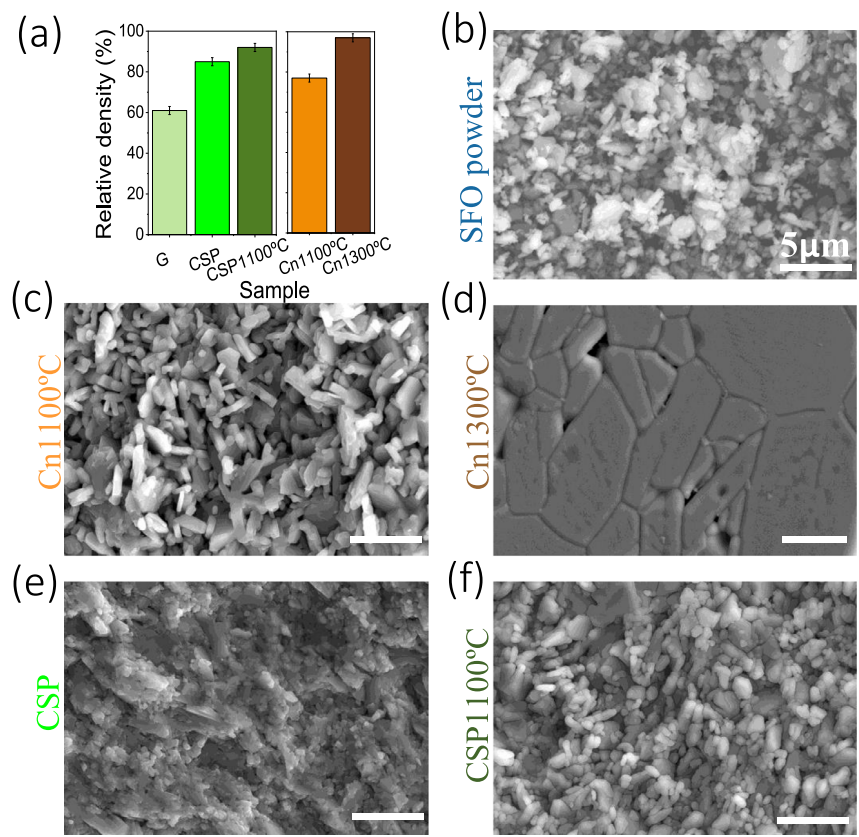

**Fig. 2.** (a) Relative densities for G (Step 3), CSP (Step 4), CSP1100°C (Step 5), Cn1100°C and Cn1300°C. FESEM images of (b) starting SFO powders and SFO ceramic processed: (c) Cn1100°C, (d) Cn1300°C, (e) CSP and (f) CSP1100°C. The white bars correspond to 5 μm.

Magnetic properties of the sintered ferrites were evaluated by a vibrating sample magnetometer (PPMS-VSM model 6000 controller, Quantum Design system) and acquired under a maximum applied magnetic field of 5 T at RT. Therefore, here the magnetization values are given at 5 T ($M_{5T}$) instead of saturation magnetization ($M_S$) [30].

## 3. Results and discussion

### 3.1. Morphological and structural properties of hexaferrite permanent magnets

The innovative strategy by CSP using GAA as a novelty plus a subsequent post-annealing was followed to sinter dense hexaferrite ceramics with improved magnetic properties. During the sintering process of the SFO, the relative density of the pieces was evaluated and data are shown in Fig. 2a. The piece obtained from the mixture of SFO and GAA compacted in Step 3 (pellet G) presents a relative density of 61%, which is increased to 85% after the CSP (Step 4). Subsequent annealing at 1100 °C for 2 h (CSP1100°C) induces a remarkable increase of the relative density to 92%, which is a satisfactory ceramic densification for many current magnetic applications. This density value supposes a significant increase of 20% with respect to the hexaferrite magnet sintered by a conventional route at same temperature for 4 h (Cn1100°C). It is required to pass the temperature from which the ferrite particle size grows significantly (Cn1300°C) to get the highest relative density of 97%, with the negative consequences on the magnetic properties.

The effect of the sintering process on the ceramic morphology, analyzed by FESEM, is shown in Fig. 2b–f. The starting SFO powders show a platelet-like morphology as commented above with a bimodal average size of 1–3 μm and 100–500 nm, which grow as the conventional route is followed at 1100 °C to more homogeneous particles with an average value of 1.7(3) μm. A particle growth with an average value of 7(2) μm is reached at 1300 °C. It should be mentioned that the presence of triple point junctions having 120 ° angles with straight particle boundaries and the existence of closed porosity indicate that the third stage of sintering is achieved at 1300 °C. This is not observed for Cn1100°C where the particle morphology continues retaining a platelet shape with a slight growth and an open porosity that point to the microstructural equilibrium not being completely attained.

When the SFO powders are processed by CSP using GAA as organic solvent, a particle refinement is identified with respect to the starting powder. The coalescence of smaller particles in equiaxed particles and an average size of 1.1(4) μm are noted. After the post-annealing process of the CSP ceramics at 1100 °C, a change in the particle morphology is identified with more rounded and homogeneous particles of around 1.0(2) μm in size that signaled the influ-ence of the sintering route in the effective particle growth inhibition.

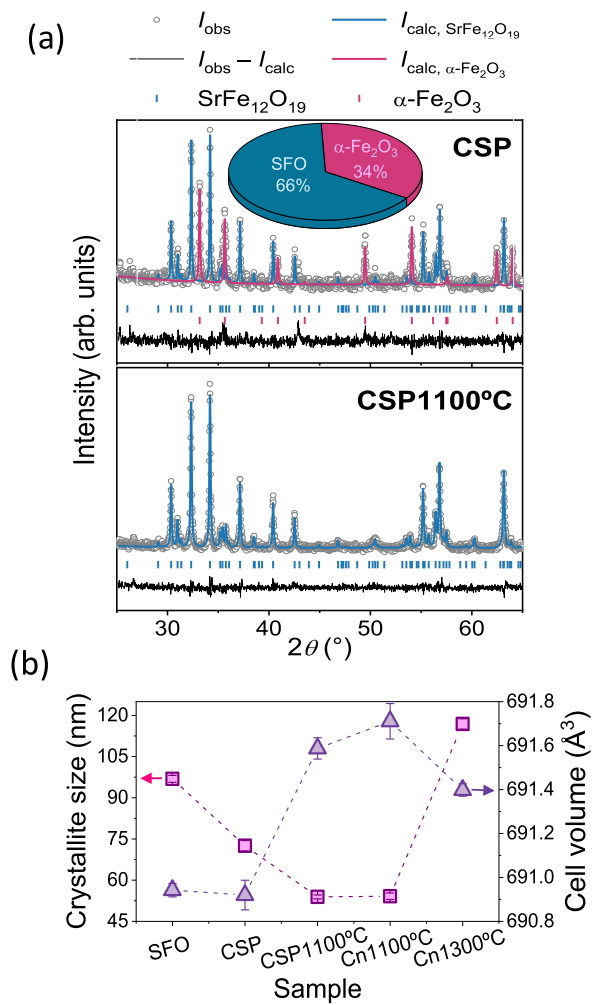

**Table 1**
Results of LCF performed in the XANES range showing the phase composition of the permanent magnets sintered. Fraction of each phase is calculated considering the results at both absorption edges.

| Sample | SFO (%) | SrO (%) | $\alpha$-Fe$_2$O$_3$ (%) |
|---|---|---|---|
| **CSP** | 46.0(5) | 28.2(4) | 25.8(5) |
| **CSP1100°C** | 92.4(3) | 7.6(3) | 0.0(1) |
| **Cn1100°C** | 96.2(3) | 3.8(5) | 0.0(1) |
| **Cn1300°C** | 99.2(4) | 0.8(1) | 0.0(1) |

**Fig. 3.** (a) Rietveld refinement (continuous lines) of the XRD patterns (grey dots) for CSP and CSP1100°C, indicating the phase proportion after the CSP. (b) The crystallite size (squares) and the unit cell volume (triangles) are also represented along the values calculated for the ceramics sintered by conventional route and the starting SFO powders.

The identification of phases and long-range structural modifications of the SFO lattice induced with the sintering process have been investigated by means of XRD. Fig. 3a shows the XRD patterns related to CSP and CSP1100°C. For the CSP magnet, along with the SFO diffraction pattern [22], Bragg's peaks related to $\alpha$-Fe$_2$O$_3$ phase [23] are recognized and a crystalline fraction of 34% is obtained by Rietveld refinement. Thus, a partial destruction of the ferrite phase is induced by the CSP under pressure, which is recovered after the annealing process at 1100 °C (CSP1100°C sample) that also helps to improve the densification.

The model-to-data agreement of the XRD patterns for the rest of samples is shown in Fig. S1 in the supplementary information (SI). In the ceramics sintered by the conventional methodology, only the reflections associated with ferrite are identified [22], and no other crystalline secondary phases are detected. In addition, the SFO lattice parameters (see SI) and unit cell volume (Fig. 3b) calculated for the CSP ceramic are similar to those obtained for the SFO starting powders, while the rest exhibits an expansion of the lattice. Concerning the crystallite size, just the Cn1300°C ceramic shows a growth with a value around 117 nm larger than the SFO powder (97 nm). It should be pointed out that all samples show large differences between the particle and crystallite sizes, indicating the large number of crystallites are forming the SFO particles. The crystalline growth in the sintered magnets during the sintering processes could have been inhibited by the presence of secondary and amorphous phases identified by XRD and XAS.

Possible non-crystalline phases, unidentified by XRD, may be given during the sintering process, especially after the CSP step. In addition, important modifications on the Fe and Sr stoichiometry and on the short-range SFO structure could have happened. This investigation was carried out by XAS experiments, at the Fe and Sr K-edge. Both X-ray absorption near-edge structure (XANES) and EXAFS experiments were performed.

XANES results for the sintered ceramic magnets along with the XANES profiles of the starting SFO, $\alpha$-Fe$_2$O$_3$ and SrO references are shown in Fig. 4a. From the absorption edge position at the XANES regions, the average valence of the absorbing atoms for all ceramics is calculated to be 3 + and 2 + for the Fe and Sr cations, respectively (see Fig. S2 in SI) [31,32]. Besides, a similar XANES profile of the sintered ferrite to that of the SFO powders is identified in both absorption edges, with changes in some resonances for the case of the CSP ferrite. By a linear combination fitting (LCF) of the XANES signal, around a 46% of SFO and a 26% of $\alpha$-Fe$_2$O$_3$ phase are obtained for the CSP ceramic (Fig. 4b and Table 1). These findings suppose a ratio of $\alpha$-Fe$_2$O$_3$ and SFO phases of 0.56, corroborating the XRD results where a $\alpha$-Fe$_2$O$_3$/SFO ratio of 0.52 was found. Interestingly, along the $\alpha$-Fe$_2$O$_3$ phase, already identified by XRD, here a proportion of a second secondary phase is recognized: SrO phase with a 28%, which may be considered an amorphous compound generated during the CSP step from the Sr cations located in the Sr ferrite structure. As the post-annealing of the CSP magnet is carried out, the ferrite phase is almost completely recovered, obtaining a slight quantity of amorphous SrO (8%) and without evidences of the hematite polymorph. By comparing the sintering process by using GAA with the standard aqueous solvent in the CSP step [15], the aqueous solvent inhibits the densification of material and the recovery of SFO after the post-annealing (see Figs. S4 and S5 in SI). During the sintering process by CSP using aqueous solutions, more problems in maintaining the formulation of the compound are identified [21,33]. Besides, a low dissolution of particle surface during the first stages of CSP can be considered taking into account the similar morphological characteristics to the starting SFO particles and the lower relative density of 84% with respect to the permanent magnets sintered by CSP using pure solvent (GAA).

Particularly, the pre-edge peak in the XANES region at the Fe K-edge (related to 1s $\rightarrow$ 3d transitions) [34] shows intensity modifications with the sintering process, see Figs. 4a and S3 in SI. The highest intensity is found for the Cn1300°C sample, indicating a distortion of site symmetry for Fe$^{3+}$ cations respect to those of the starting SFO powders that show a lower intensity with a coordination closer to the octahedron. In addition, for the Cn1300°C sample, the intensity at the whiteline is the lowest in agreement with the largest decrease of the coordination number as the EXAFS results confirm below. With respect to the position of the pre-edge peak, similar values are identified for all samples, indicating that the oxidation state of Fe cations is the same, 3 + [31], as calculated above.

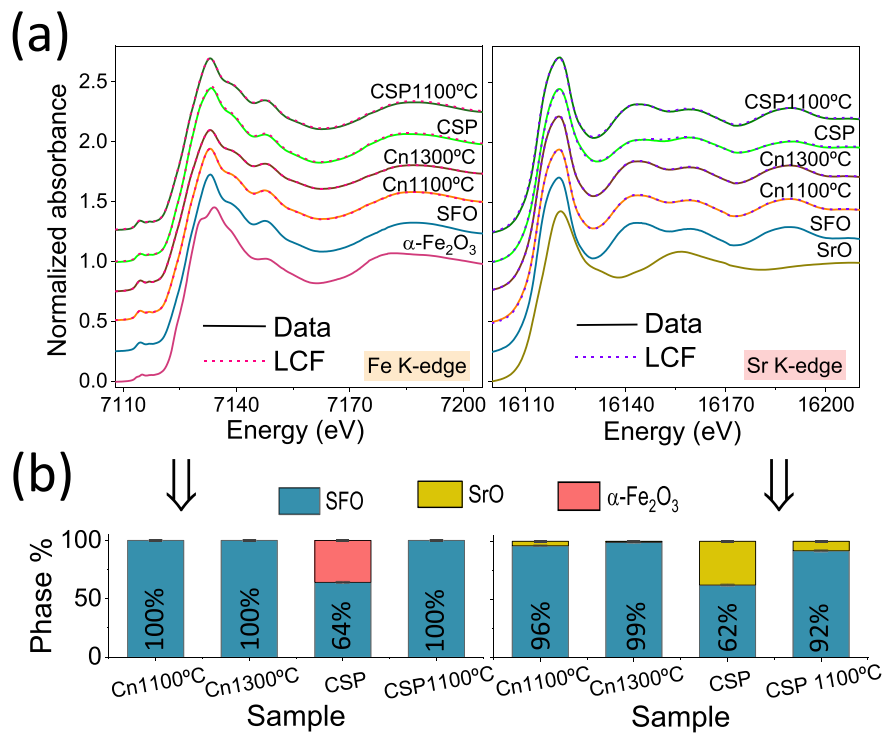

Fig. 4. (a) XANES spectra (continuous lines) at the Fe and Sr K-edge and LCF curves (dashed lines) for CSP, CSP1100°C, Cn1100°C and Cn1300°C. (b) Fraction of each phase obtained from the LCF of the XANES spectra at the Fe K-edge (left) and the Sr K-edge (right) for all ceramics sintered.

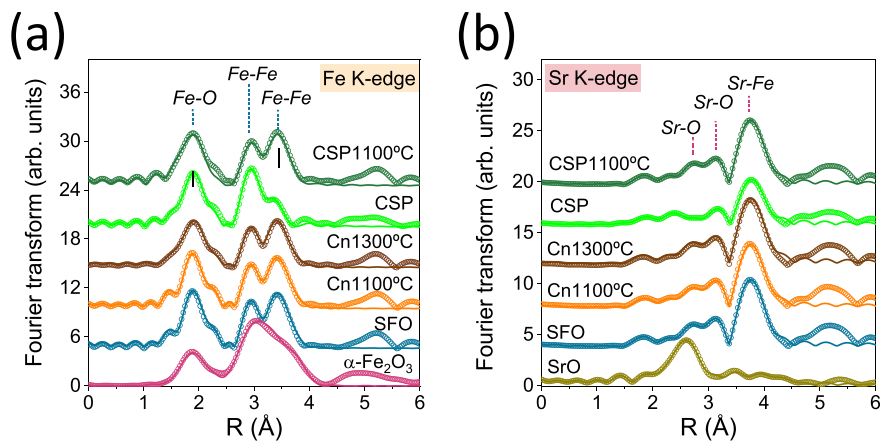

Fig. 5. Fourier transform modulus of the EXAFS spectra (dots) at the (a) Fe and (b) Sr K-edge and best-fitting simulations by a three-shell model (continuous lines) of the processed magnets and SFO powders, along with $\alpha$-$Fe_2O_3$ and SrO references.

Short-range ordering analysis of cations around the Fe and Sr cations and the parameters obtained by EXAFS are shown in Fig. 5 and Table 2. The most significant alterations are found in the structure of the SFO-based ceramic after the CSP. Some modifications at the second and the third Fe-Fe shells are distinguished close to the signal of the hematite structure, corroborating the phase transformation with the CSP and the XANES data. With respect to the Sr ions, an intensity reduction at the three modeled coordination shells is obtained after the CSP (Table 2). As the post-annealing is performed, the SFO structure is recovered achieving similar EXAFS parameters than the starting SFO powders. For all samples, no modifications in the DW factors are observed at both absorption edges with respect to the reference powders, neither with the sintering step nor with the sintering process. To conclude, the reduction of the coordination number at both edges for the sintered ceramics (mostly in Cn1300°C) with respect to the starting SFO powders (see Table 2) should be mentioned, which indicates the loss of some atoms into the structure inducing defects (e.g. vacancies) during the sintering process [14].

Hence, the selectivity of XAS analysis has allowed to detect what happens to the ferrite during the sintering process and where the Sr cations are located as the ferrite structure is transformed (i.e. forming a SrO amorphous secondary phase). Furthermore, these findings can explain why Sr surplus is generally required to produce $\alpha$-$Fe_2O_3$-free SFO as identified in some studies [35,36], which has represented an issue in ferrite manufacturing for a long time. While the stoichiometric Sr:Fe ratio of the ferrite is 1:12 [3], ratios as high as 1:1 have been needed for its synthesis, which entails a great Sr excess resulting in the formation of amorphous SrO.

Table 2
Results of the EXAFS fittings by a three-shell model at the Fe and Sr K-edge of ferrite ceramics. EXAFS parameters are compared with results obtained from magnets processed by the conventional route and starting SFO, $\alpha$-Fe$_2$O$_3$ and SrO references.

| Sample | Fe K-edge | | | | Sr K-edge | | | |
|---|---|---|---|---|---|---|---|---|
| | shell | N | R (Å) | DW(Å$^2$) | shell | N | R (Å) | DW(Å$^2$) |
| **Starting SFO powder** | Fe-O | 6 | 1.960(4) | 0.013(2) | Sr-O | 6 | 2.795(3) | 0.007(2) |
| | Fe-Fe | 6 | 2.970(6) | 0.006(2) | Sr-O | 6 | 2.99(1) | 0.004(2) |
| | Fe-Fe | 6 | 3.465(4) | 0.009(1) | Sr-Fe | 15 | 3.690(6) | 0.015(2) |
| **Cn1100°C** | Fe-O | 5.8(2) | 1.959(4) | 0.013(1) | Sr-O | 5.4(4) | 2.79(1) | 0.007(2) |
| | Fe-Fe | 4.7(1) | 2.968(8) | 0.005(2) | Sr-O | 5.5(3) | 2.98(1) | 0.004(1) |
| | Fe-Fe | 6.1(2) | 3.459(4) | 0.009(1) | Sr-Fe | 14.3(3) | 3.692(7) | 0.015(2) |
| **Cn1300°C** | Fe-O | 5.3(1) | 1.954(4) | 0.013(1) | Sr-O | 5.6(4) | 2.792(9) | 0.006(2) |
| | Fe-Fe | 3.9(2) | 2.956(8) | 0.005(1) | Sr-O | 5.9(3) | 2.98(1) | 0.004(2) |
| | Fe-Fe | 6.1(1) | 3.451(4) | 0.009(1) | Sr-Fe | 14.8(1) | 3.688(6) | 0.015(1) |
| **CSP** | Fe-O | 5.8(1) | 1.959(5) | 0.012(1) | Sr-O | 3.8(3) | 2.79(1) | 0.006(2) |
| | Fe-Fe | 5.3(1) | 2.961(8) | 0.003(2) | Sr-O | 4.5(3) | 2.98(1) | 0.005(2) |
| | Fe-Fe | 3.6(2) | 3.417(8) | 0.005(2) | Sr-Fe | 9.5(1) | 3.69(1) | 0.015(1) |
| **CSP1100°C** | Fe-O | 6.1(2) | 1.956(4) | 0.012(1) | Sr-O | 6.0(4) | 2.79(1) | 0.006(2) |
| | Fe-Fe | 5.8(2) | 2.968(8) | 0.006(1) | Sr-O | 5.8(3) | 2.98(1) | 0.004(2) |
| | Fe-Fe | 6.0(2) | 3.452(4) | 0.008(1) | Sr-Fe | 15.0(3) | 3.691(6) | 0.015(2) |
| **$\alpha$-Fe$_2$O$_3$ powder reference** | Fe-O | 6 | 1.96(1) | 0.013(1) | | | | |
| | Fe-Fe | 4 | 2.948(7) | 0.006(1) | | | | |
| | Fe-Fe | 3 | 3.375(8) | 0.006(1) | | | | |
| **SrO powder reference** | | | | | Sr-O | 6 | 2.621(3) | 0.011(1) |
| | | | | | Sr-Sr | 12 | 3.728(6) | 0.021(2) |

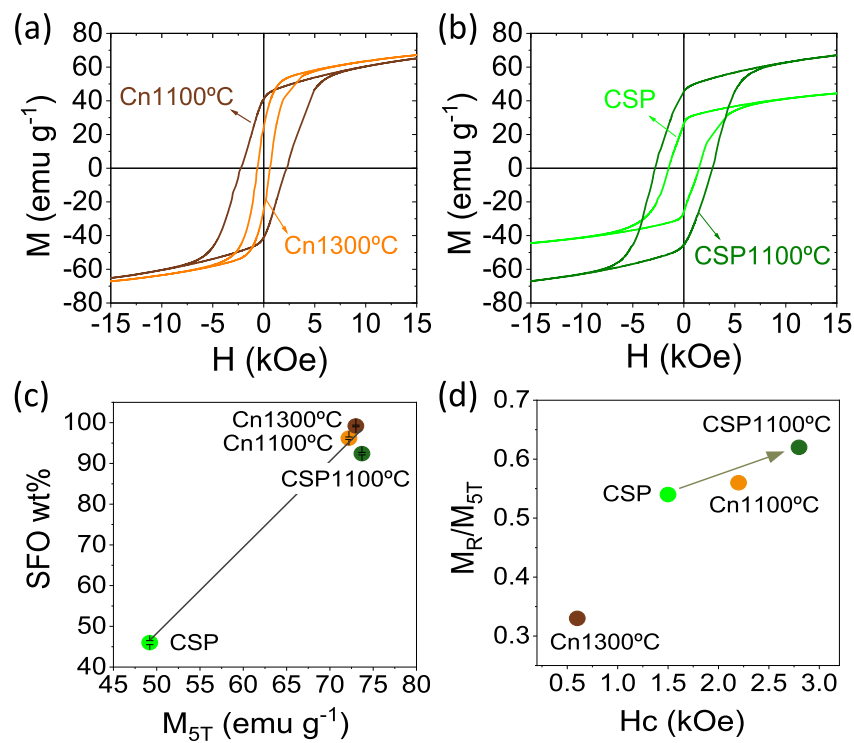

**Fig. 6.** Hysteresis loops representing the magnetic response for (a) Cn1100°C and Cn1300°C and (b) CSP and CSP1100°C. (c) Dependence between the M$_{5T}$ values and SFO wt%. (d) M$_R$/M$_{5T}$ ratio versus H$_C$ obtained from the magnetic curves.

### 3.2. Magnetic performance of hexaferrite permanent magnets: improving functional response

The functional magnetic response of the sintered ferrites was evaluated and represented in Fig. 6. For all ceramic magnets a characteristic hysteresis loop of a ferromagnetic material is measured with singular properties depending on the sintering process. Cn1100°C presents a H$_C$ of 2.2 kOe and a M$_{5T}$ of 72.2 emu g$^{-1}$. However, despite the interesting magnetic properties as hard magnetic material, the relative density of the magnet is low and the mechanical integrity is poor. By increasing the sintering temperature to 1300 °C, densification is achieved as shown above, but reducing considerably the H$_C$ to 1.0 kOe due to the particle growth (see Fig. 6a). Conversely, for CSP magnet (see Fig. 6b), who presents 85% relative density, the partial decomposition of SFO into $\alpha$-Fe$_2$O$_3$ and SrO compounds with the sintering process is traduced in a reduction of coercivity (H$_C$ of 1.5 kOe) and magnetization at 5 T (M$_{5T}$ = 49.2 emu g$^{-1}$). These results are in agreement with the presence of non-magnetic SrO phase and the low saturation magnetization of $\alpha$-$Fe_2O_3$. However, after the post-annealing at 1100 °C (CSP1100°C), in addition to the increased density to 92%, a clear improvement of magnetic properties is obtained with a $H_C$ of 2.8 kOe and $M_{5T}$ = 73.7 emu g$^{-1}$, as a consequence of recrystallization of the ferromagnetic SFO phase. For that, the role of GAA in the CSP is crucial allowing also to refine the particle size during the sintering process, which results in an improvement of the coercivity value. In all cases, the $BH_{max}$ values are estimated from the open hysteresis loops and non-orientated samples, obtaining the highest merit figure for the CSP1100°C magnet (see Table S3 in SI). Besides, the dependence between the $M_{5T}$ values and the wt% of the ferromagnetic phase (i.e. SFO) is also recognized and displayed in Fig. 6c. These properties are perfectly competitive when compared to commercial ferrite magnets (Hitachi metals NMF-7C series offer $H_C$ = 2.8–3.3 kOe and $M_S$ = 68 emu g$^{-1}$) [37]. Additionally, given that competent ferrite magnets can be sintered at 1200–1250 °C by the incorporation of secondary phases, our procedure allows reaching similar performance at 8–12% lower temperatures which means an energy efficiency of at least 25%.

During the CSP, a slight increment of the degree of magnetic alignment between particles is also attained, with a $M_R/M_{5T}$ ratio of 0.54, exceeding the predicted value for an assembly of non-interacting particles randomly oriented [38]. The $M_R/M_{5T}$ ratio rises even more up to 0.62 with the post-annealing for the CSP1100°C magnet, as Fig. 6d shows. This magnetic improvement is likely due to a favourable arrangement under pressure of the ferrite platelets facilitated by the liquid solvent.

## 4. Conclusion

Herein a strategy to sinter ceramic magnets based on Sr-hexaferrites with interesting structural and magnetic characteristics is developed. An approach combining the CSP as an intermediate step, in which GAA is employed as pure solvent, plus a subsequent post-annealing has densified the ceramic magnets to a relative density of 92%. Despite the partial transformation of the magnetic phase during the CSP, which is accompanied by a reduction of the magnetic properties, after the post-annealing of the CSP ceramic at 1100 °C the magnetic phase is recovered with particle refinement. An improvement in the final functional response of the ceramic magnets is achieved, greatly surpassing the performance of the hexaferrite magnets processed by conventional route and with very competitive response at lower temperatures. These investigations are the first step in a very promising direction, demonstrating a feasible and green sintering way towards the development of dense rare-earth free oxide permanent magnets for a large number of technological applications.

## Declaration of Competing Interest

The authors declare that they have no known competing financial interests or personal relationships that could have appeared to influence the work reported in this paper.

## Acknowledgments

This work has been supported by the Ministerio Español de Ciencia, Innovación y Universidades (MCIU) through the projects MAT2017-86540-C4-1-R and RTI2018-095303-A-C52 and The European Commission through the AMPHIBIAN Project (720853). Part of these experiments was performed at the CLÆSS beamline at ALBA Synchrotron with the collaboration of ALBA staff. A.S. acknowledges financial support from Comunidad de Madrid for an "Atracción de Talento Investigador" Contract (2017-t2/IND5395). C.G.-M. and A.Q. acknowledge financial support from Ministerio Español de Ciencia e Innovación (MICINN) through the "Juan de la Cierva" Program (FJC2018-035532-I) and the "Ramón y Cajal" Contract (RYC-2017-23320).

## Supplementary materials

Supplementary material associated with this article can be found, in the online version, at doi:10.1016/j.actamat.2021.117262.